# Summary of 2nd International Workshop on Requirements Engineering and Testing (RET 2015)
## Co-located with ICSE 2015

Elizabeth Bjarnason, Markus Borg
Lund University
Sweden
{1st.Last}@cs.lth.se

Mirko Morandini
Fondazione Bruno Kessler
Italy
morandini@fbk.eu

Michael Unterkalmsteiner
Blekinge Inst. of Technology
Sweden
mun@bth.se

Michael Felderer
University of Innsbruck, Austria
michael.felderer@uibk.ac.at

Matthew Staats
Google Inc., Switzerland
staatsm@gmail.com

## ABSTRACT
The RET (Requirements Engineering and Testing) workshop series provides a meeting point for researchers and practitioners from the two separate fields of Requirements Engineering (RE) and Testing. The goal is to improve the connection and alignment of these two areas through an exchange of ideas, challenges, practices, experiences and results. The long term aim is to build a community and a body of knowledge within the intersection of RE and Testing, i.e. RET. The 2nd workshop was held in co-location with ICSE 2015 in Florence, Italy. The workshop continued in the same interactive vein as the 1st one and included a keynote, paper presentations with ample time for discussions, and a group exercise. For true impact and relevance this cross-cutting area requires contribution from both RE and Testing, and from both researchers and practitioners. A range of papers were presented from short experience papers to full research papers that cover connections between the two fields. One of the main outputs of the 2nd workshop was a categorization of the presented workshop papers according to an initial definition of the area of RET which identifies the aspects RE, Testing and coordination effect.

### Categories and Subject Descriptors
D.2.1 [**Requirements/Specifications**], D.2.4 [**Software/Program Verification**] D.2.5 [**Testing and Debugging**]

## 1. OBJECTIVES
The main objective of this 2nd Requirements Engineering and testing (RET) workshop[1] was to establish a common understanding of what the interaction of Requirements Engineering (RE) and Testing, i.e. RET, entails; to share and discuss challenges and potential solutions to manage and improve this interaction. The workshop provided a forum for exchanging ideas and best practices for aligning RE and testing, with the potential of collaboration on this topic both between academics and between industry and academia. Towards this, RET invited submissions exploring how to coordinate RE and testing including processes, practices, artefacts, methods, techniques, and tools. Submissions on softer aspects like the communication between roles in the engineering process were also welcomed.

RET accepted technical papers with a maximum length of 8 pages presenting research results or industrial practices related to the coordination of Requirements Engineering and Testing, as well as position papers with a minimum length of 2 pages introducing challenges, visions, positions or preliminary results within the scope of the workshop. Experience reports and papers on open challenges in industry were especially welcome.

## 2. ORGANIZATION
The 2nd International Workshop on Requirements Engineering and Testing (RET'15) was held on May 18, 2015, and was co-located with the 37th International Conference on Software Engineering (ICSE 2015). The website of the workshop is available online at http://ret.cs.lth.se/15/. The workshop was organized by Elizabeth Bjarnason (Lund University) as a general chair, and Mirko Morandini (University of Trento) and Markus Borg (Lund University) as program co-chairs, as well as, Michael Unterkalmsteiner (Blekinge Institute of Technology), Michael Felderer (University of Innsbruck) and Matt Staats (Google Zurich) as co-chairs.

## 3. PROGRAM SUMMARY
The program of RET'15 comprised an introductory part, three paper presentation sessions and an interactive exercise. Besides a welcome note presenting the main objectives of the workshop, Professor Mats Heimdahl gave an invited talk "Requirements and Tests: When Do We Have Enough?" where he discussed the need for requirements coverage (test coverage of requirements) rather than merely test coverage of code. While code coverage is used to measure the quality of tests, requirements coverage gives an indication of how well the software measures up to the requirements. Thus requirements coverage measures could be used to indicate how well the expectations of the customers and users are fulfilled, i.e. quality of the software. But how should we measure this kind of test coverage? One test case per requirement is probably not enough. Tests can be generated from code to ensure 100% MCDC (modified condition/decision coverage), but what does this really mean? Mats presented two cases: one where 95% of requirements were covered and 60% of the source code, the other where 60% of the requirements were covered and 95% of the source code. We need to understand more about what these different coverage criteria mean in practice and how to ensure and measure good requirements coverage.

The keynote was followed by a presentation by the workshop chair of the current state of the conceptual map of RET. The initial version of this map was produced at the 1st International Workshop on Requirements Engineering and Testing (RET2014[2]) (co-located with the 22nd International Requirements Engineering Conference in Karlskrona, Sweden, 2014). Since then the organisers have revised and grouped the initial 20 topics into 7 main topic areas and 18 topics. In addition, an initial definition of RET has been formulated. Both of these were further discussed and refined during the exercise in the afternoon. See Section 4 for more details.

The first paper session comprised three position papers on aspects of quality related to RET. The talk "Aligning Quality Requirements and Test Results with QUPER's Roadmap View for Improved High-Level Decision-Making" (A1) proposed showing test case results for quality requirements alongside the various requirements-related values in the road map view of the QUPER model, e.g. levels for competitors, utility and differentiation breakpoints. An initial evaluation indicates that practitioners find this approach valuable in making more informed decisions about the implementation of quality requirements.

---

[1] http://ret.cs.lth.se/15

[2] http://webhotel.bth.se/re14/ret/

The talk "Requirement-Centric Reactive Testing for Safety-Related Automotive Software" (A2) addressed the issue of testing the many and complex interactions of the constantly increasing number of software functions in software systems with its environment through reactive testing. The paper proposes a taxonomy of reactive testing including factors such as coverage and input/output values of reactive tests. This taxonomy can be used by testers to explore and identify which types of test reactivity the testing should have and thereby support them in better aligning the testing to the requirements. The taxonomy is also seen to improve the communication between testers and towards other stakeholders thereby improving the coordination between requirements and testing.

The talk "Play-Testing and Requirements Engineering: Implications for Research and Teaching" (A3) presented a case study on how play testing was used to validate the game World of Warcraft. The study aims to better understand this testing which is often performed by users without any RE or testing competence and unveils that those who perform play-testing are usually people originating from the development engineers, i.e. colleagues, family, friends, friends of friends. Furthermore there is a tight connection between play testing and revising the requirements based on the play-tester's feedback, thus gradually improving the requirements through iterative test-requirements loops.

The second paper session comprised two full technical papers on research and practice. In the talk "It's the Activities, Stupid! A New Perspective on RE Quality" (B1) it was suggested that RE quality is context-specific, i.e. depends on what the requirements artefacts are to be used for rather than the (stand-alone) quality of the SRS. A methodology and an initial validation of defining RE artefact quality using an activity-based model was presented.

In the talk "Weekly Round Trips from Norms to Requirements and Tests: an Industrial Experience Report" (B2) a case study was presented where an iterative approach to gradually re-engineering a legacy system was studied. The re-engineering project was largely dependent on transforming regulatory norms into requirements, before commencing implementation and testing. The identified challenges include communication between the heterogeneous teams (experts in normative rules, RE, development and testing), handling automated testing, combining agility with control, and managing resource and time estimation.

The third paper session comprised three papers on envisioned solutions; one full technical paper and two position papers. The first talk on "Configuring Latent Semantic Indexing for Requirements Tracing" (C1) discussed the challenge of configuring the algorithms used for automatically deriving traceability links, e.g. between requirements and test artefacts. The evaluation of a fully automated technique for determining LSI configurations was presented. This technique relies only on heuristic metrics calculated on the artefacts, and does not require the input of experts or known links thereby reducing the time and cost of configuring the trace recovery.

The talk "Towards Automatic Constraints Elicitation in Pair-Wise Testing Based on a Linguistic Approach – Elicitation Support Using Coupling Strength" (C2) presented a technique for supporting combinatorial test design based on the elicitation of combinations of constraints from the requirements specification. By deriving information about constraints using a measurement of distance between parameters representing the 'coupling strength' the test space can be reduce by avoiding non-relevant combinations.

In the talk "Visual Requirements Specification and Automated Test Generation for Digital Applications" (C3) the challenge of testing visual requirements was discussed. A technique was proposed for generating test cases based on the information stored in a visual (prototype) requirements specification. A tool implementing this technique is currently being evaluated.

In the second half of the afternoon an interactive exercise was performed. This was opened with an overview of how little the workshop papers share references, thus showing how diverse the current RET community is and indicating the need for a joint understanding of what constitutes RET. During the interactive exercise the participants discussed how to define RET and validated the RET map (outcome of RET'14). The exercise and its outcome is presented in the next section.

## 4. DEFINING THE AREA OF RE AND TESTING

The main aim of the interactive exercise was to jointly define and further shape and map the area of requirements engineering and testing, i.e. RET. The participants were divided into 3 groups and given approx. 1 hour to discuss before the exercise was summarised in plenum.

The participants were provided with a map of the area of RE and Testing and an initial definition of RET. The map consisted of 6 main topic areas and 18 topics in total (see Appendix). These were derived from the output of the mapping exercise performed at RET'14 and refined during the intermittent period by the workshop organizers. In addition, an initial version of the RET definition was presented, namely

- RET covers activities which adjust or arrange the efforts involved in Requirements Engineering and/or in Testing in such a way as to support the coordination between these efforts.

- RET can either address both RE and Testing, or one of these two areas with an effect on the other.

- RET addresses the activities of RE and Testing, rather than merely considering their input/output.

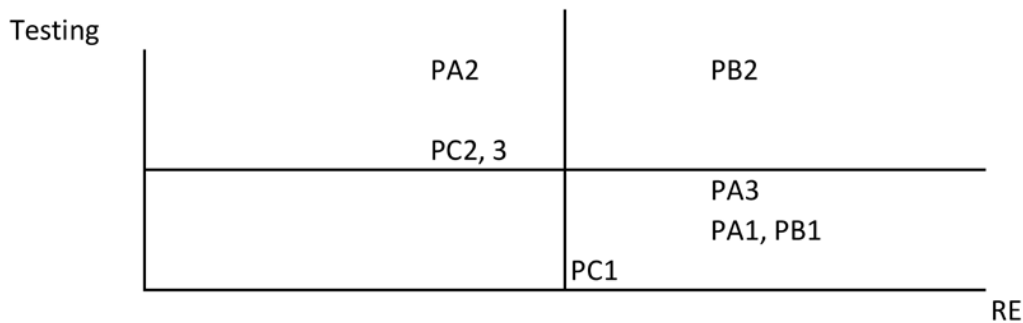

**Figure 1. The joint categorization of the RET'15 papers into a RET grid, showing the spread of the RE vs Testing emphasis presented**

The assignment given to each group consisted of analysing the RET'15 workshop papers in order to identify a) the RET topic (relative the RET map) and b) the RE, testing and coordination aspects (compare with RET definition) of each paper. In addition, each paper was placed on a grid with axes RE and testing (the RET grid) in order to investigate the degree of RE vs testing aspects in the current RET papers. Finally, the participants listed their own RET-related articles and analysed them in the same way, i.e. relative the RET map and the initial version of the RET definition. The outcome of this group activity is reporting in the appendix.

Summing up the placement of the workshop papers in plenum gave rise to some disagreements of what the RE, the testing and the coordination aspects are of the individual papers, indicating the importance of group reflections on this topic.

The main outcome of the exercise was to initiate reflections around RET. In particular, the RET map seems to have stabilized although it was suggested to add a topic to cover aligning RE and testing at various levels within the V-model.

The RET definition generated several discussion points. Even though it is a good start it needs to be refined and clarified in order to be a workable definition of the area. Specifically it was unclear to the participants what constituted the actual definition and what was clarifying phrases, and how these are to be combined (with and or or), i.e. the current version is ambiguous. Furthermore, the use of this definition needs to be considered, the organisers intention is not to unduly restrict the research covered by the RET workshop, but rather to clarify the focus of the workshop.

## 5. FUTURE
We plan to organise the workshop again next year since the topic is very relevant both to industry and within academia, as seen by the interest and positive response to this year's RET workshop from participants and PC members. Our aim is to organise RET'16 co-located with REFSQ 2016 (http://refsq.org/2016/). Due to high industrial interest in this topic REFSQ specifically elicits workshops on RE and testing. Therefore we believe the industrial participation can be increased for RET'16 by co-locating with REFSQ. If the workshop is accepted the date for paper submissions will be January 9, 2016, with author notification on January 30, 2016.

## 6. ACKNOWLEDGEMENTS

We want to thank the participants of the workshop and all the authors of submitted papers for their important contribution to the event. In addition, we want to thank the organizers of the 37th International Conference in Software Engineering (ICSE'15) and the members of the program committee listed below:
  Armin Beer, Beer Test Consulting, Austria
  Ruzanna Chitchyan, Leicester University, UK
  Nelly Condori-Fernandez, PROS Research Centre, Spain
  Robert Feldt, Blekinge Institute of Technology, Sweden
  Vahid Garousi, University of Calgary, Canada
  Joel Greenyer, University of Hannover, Germany
  Mats Heimdahl, University of Minnesota, USA
  Andrea Herrmann, Herrmann & Ehrlich, Germany
  Jacob Larsson, Capgemini, Sweden
  Per Lenberg, SAAB ATM Sensis AB, Sweden
  Emmanuel Letier, University College London, UK
  Annabella Loconsole, Malmö University, Sweden.
  Alessandro Marchetto, FIAT research, Italy
  Cu Duy Nguyen, University of Luxembourg, Luxembourg
  Magnus C. Ohlsson, System Verification, Sweden
  Barbara Paech, University of Heidelberg, Germany
  Anna Perini, FBK, Trento, Italy
  Dietmar Pfahl, University of Tartu, Estonia
  Giedre Sabaliauskaite, Singapore Univ. of Techn. & Design
  Kristian Sandahl, Linköping University, Sweden
  Hema Srikanth, IBM, USA
  Paolo Tonella, Fondazione Bruno Kessler, Italy
  Marc-Florian Wendland, Fraunhofer FOKUS, Germany
  Magnus Wilson, Ericsson, Sweden
  Krzysztof Wnuk, Lund University, Sweden
  Yuanyuan Zhang, University College London, UK

We also want to thank our steering committee for their support:
  Jane Cleland-Huang, DePaul University, USA
  Mats Heimdahl, University of Minnesota, USA
  Jane Huffman Hayes, University of Kentucky, USA
  Marjo Kauppinen, Aalto University, Finland
  Per Runeson, Lund University, Sweden
  Paolo Tonella, Fondazione Bruno Kessler, Italy

## 7. APPENDIX: EXERCISE MATERIAL
The RET map on which the exercise was based consisted of the following topics:
1. **Processes and practices**
1.1. Processes and practices
1.2. RET at different lifecycle phases
1.3. Organizational aspects, including education
1.4. Collaboration and communication
1.5. Specific contexts, e.g. distributed development, OSS, continuous deployment, safety-critical
2. **Requirements-based testing**
2.1. Reqts-based testing processes & techniques
2.2. Model-based testing
2.3. TDD and BDD
3. **Quality of requirements**
3.1. Reqts quality impact on test quality, reqts testability
3.2. Reqts validation through testing
4. **Maintaining RET alignment**
4.1. Managing change/evolution
4.2. Traceability (requirements – test cases) incl regression testing
4.3. RET metrics
5. **RET trade offs**
5.1. Scalability including big data
5.2. "Good enough" RET, effort vs ROI
6. **Quality requirements**
6.1. RET alignment for QR/NFR
7. **Tools & techniques**
7.1. Automation techniques
7.2. Tools for monitoring and managing RET alignment
7.2 Initial RET Definition

The workshop participants were presented with the following initial version of a definition of the area of RET, as defined by the workshop organizers:

- RET covers activities which adjust or arrange the efforts involved in Requirements Engineering and/or in Testing in such a way as to support the coordination between these efforts.

- RET can either address both RE and Testing, or one of these two areas with an effect on the other.

- RET addresses the activities of RE and Testing, rather than merely considering their input/output.

Each group categorized the RET15 workshop papers according to the RET map and the initial RET definition. The RE, Testing and coordination aspect for each paper was identified and the paper was placed on a RET grid consisting of one axis for degree of RE and one axis for degree of Testing. The resulting RET grid for each group is found in Figure 2.

| | Papers | RET topic | RE aspect | dgr | Testing aspect | dgr | Effect | RET topic | RE aspect | dgr | Testing aspect | dgr | Effect | RET topic | RE aspect | dgr | Testing aspect | dgr | Effect |
|---|---|---|---|---|---|---|---|---|---|---|---|---|---|---|---|---|---|---|---|
| A1 | Berntsson Svensson et al. | Aligning Quality Requirements and Test Results with QUPER's Roadmap View for Improved High-Level Decision-Making | 2 | PM of QRs | 4 | output from testing | 2 | Improved decision support | 6 | Understand where to put reqts target | 5 | Prioritisation of test | 3 | trace of test reqsults to requirements, RE knows whether reqts are fulfilled | 6.1, 5.2 | quality targets | 6 | testing quality | 2 | understanding the quality targets |
| A2 | Mjeda, Hinchey | Requirement-Centric Reactive Testing for Safety-Related Automative Software | 6.1 | Elicitation of reqts | 2 | MBT, Test design, standards compliants, testing | 5 | On test design | | | | | | | 2.1 | none | 1 | taxonomy of reactive testing | 5 | support test creation and communication |
| A3 | Daneva | Play-Testing and Requirements Engineering: Implications for Research and Teaching | 6.1 | Reqts validation | 6 | Usability, playability & experience testing | 3 | Requirements validation | | | | | | | 1.1, 1.5 | possible impact of play testing on RE | 2 | understanding of play-testing practices | 2 | adding game perspective fundamental knowledge (no solution proposal) |
| B1 | Femmer et al. | It's Activities, Stupid! A New Perspective on RE Quality | 3 | quality of reqts documentation | 6 | validation | 2 | improved quality of reqts doc | | | | | | | 3.1, 5.2 | quality of RE artefacts (model) | 6 | none | 1 | defining and validating RE quality |
| B2 | Tonella, Tiella | Weekly Round Trips from Norms to Requirements and Tests: an Industrial Experience Report | 1.1 | Reqts elicitation | 4 | Deriving test cases | 4 | Improved reqts-tc alignment | | | | | | | 1.1, 2.1, 1.3, 1.4, 4.2, 7.1 | Reqts-based dev and testing process | 4 | Tests derived from UI model, use cases | 4 | reporting the validity of combined agile and V-model |
| C1 | Eder et al. | Configuring Latent Semantic Indexing for Requirements Tracing | 3 | Traces between reqts | 4 | none | 1 | Improvement of traces between reqs and test artefacts | 4, 7 | traces to test cases | 3 | traces to requirements | 3 | automatically derive traces which then leads to less communication | 4.2, 7.1 | main trace artefact | 3 | traceable artefac | 2 | easier tuning of automated tracing |
| C2 | Nakagawa, Tsychiya | Towards Automatic Constraints Elicitation in Pair-Wise Testing Based on a Linguistic Approach: Elicitation Support Using Coupling Strength | 7 | Eliciting constraints from reqts doc | 3 | pair-wise testing | 5 | improve pair-wise testing with constraints elicitation | 2, 7 | Extracting constraints | 3 | Optimizing/reducing the set of test cases | 5 | Derive better constraints for testing using the requirements | 7.1, 2.1 | Source of constraints | 2 | Improve pair-wise testing | 4 | to help the tester find conditions/constraints based on the requirements |
| C3 | Singi et al. | Visual Requirements Specification and Automated Test Generation for Digital Applications | 7 | visual reqts | 5 | test case generation | 4 | achieve agile reqts and testing phases | 2 | uses requirements | 1 | test automation | 5 | generate requirements-based test cases, enables testers to test more efficiently / better coverage | 2.2, 1.5, 7.1 | Formalizing visual reqts to generate test cases | 4 | generate test cases from visual reqts | 4 | automated generation of test cases from visual reqts |

**Figure 2. The categorization of RET'15 papers per group.**